\documentclass[prb,twocolumn,aps,superscriptaddress,showpacs]{revtex4}

\usepackage{graphicx}
\usepackage{psfrag}
\usepackage{amsmath}

\newcommand{\be}{\begin{equation}}
\newcommand{\ee}{\end{equation}}
\newcommand{\bea}{\begin{eqnarray}}
\newcommand{\eea}{\end{eqnarray}}

\newcommand{\p}{\partial}

\newcommand{\la}{\langle}
\newcommand{\ra}{\rangle}

\newcommand{\ham}{{\cal H}}

\def\nn{\nonumber\\}

\begin{document}
\unitlength=1mm

\title{Lifshitz transitions and crystallization of fully-polarised dipolar Fermions in an anisotropic 2D lattice}

\author{Sam T. Carr}
\affiliation{Institut f\"ur Theorie der Kondensierten Materie, Karlsruher Institut f\"ur Technologie, 76128 Karlsruhe, Germany}
\affiliation{DFG Center for Functional Nanostructures, Karlsruher Institut f\"ur Technologie, 76128 Karlsruhe, Germany}

\author{Jorge Quintanilla}
\affiliation{School of Physical Sciences, University of Kent, Canterbury CT2 7NH, U.K.}
\affiliation{ISIS spallation facility, STFC Rutherford Appleton Laboratory, Harwell
Science and Innovation Campus, OX11 0QX, U.K.}

\author{Joseph J. Betouras}
\affiliation{Department of Physics, Loughborough University, Loughborough LE11 3TU, U.K.}

\date{\today}


\begin{abstract}
We consider a two dimensional model of non-interacting chains of spinless fermions weakly coupled via a small inter-chain hopping and a repulsive inter-chain interaction.  The phase diagram of this model has a  surprising feature: an abrupt change in the Fermi surface as the interaction is increased.  We study in detail this meta-nematic transition, and show that the well-known $2\frac{1}{2}-$order Lifshitz transition is the critical endpoint of this first order quantum phase transition.  Furthermore, in the vicinity of the endpoint, the order parameter has a non-perturbative BCS-like form.  We also study a competing crystallization transition in this model, and derive the full phase diagram.  This physics can be demonstrated experimentally in dipolar ultra-cold atomic or molecular gases. In the presence of a harmonic trap, it manifests itself as a sharp jump in the density profile.
\end{abstract}

\maketitle

\section{Introduction}

The study of phenomena which deform the Fermi surface in both electron and cold atom systems has gained much popularity recently.  For example, it has been sugested that the Fermi surface shape-changing Pomeranchuk instability\cite{Pom58,Yamase-Kohno-2000,Halboth-Metzner-2000,Oganesyan-Kivelson-Fradkin-2001,Valenzuela-Vozmediano-2001} may  describe experiments in heavy fermions,\cite{Settai-et-al-2005,Sugawara-et-al-2002,Paschen-et-al-2004}  quantum hall devices,\cite{Cooper-et-al-2002} and ruthenates,\cite{Grigera-et-al-2004,Mercure-et-al-2009}
leading to a plethora of theoretical papers on the subject.\cite{Quintanilla-Schofield-2006,Varma-Zhu-2006,Varma-Zhu-2007,Doan-Manausakis-2007,Yamase-2009a,Yamase-2009b,Lee-Wu-2009,Zacharias-Wolfle-Garst-2009,Puetter-Rau-Kee-2010}

The Pomeranchuk instability, which breaks rotational symmetry but no translational symmetries, is essentially a transition to an electronic nematic phase.\cite{Yamase-Kohno-2000,Halboth-Metzner-2000,Oganesyan-Kivelson-Fradkin-2001,Valenzuela-Vozmediano-2001} An analogue in spin systems has also been studied.\cite{Shannon-Momoi-Sindzingre-2006}  Incorporating other electronic analogues of liquid crystal phases into this picture has been put forward as a general picture of strong correlations\cite{Kivelson-Fradkin-Emery-98,Fradkin-Kivelson-Oganesyan-07} with evidence for smectic phases being observed experimentally in manganites\cite{Chen-Cheong-1996} and cuprates.\cite{Tranquada-et-al-1995}

These electronic liquid crystal phases also have a strong relation to dimensional crossover phenomena, where one can ask the question whether an array of one-dimensional chains (Luttinger liquids) coupled by a weak inter-chain hopping $t_\perp$ remains strictly one-dimensional (confinement),\cite{Clarke-et-al-94} or becomes a quasi-one-dimensional Fermi liquid (deconfinement).  Indeed, calculation methods such as self-consistent perturbation theory\cite{Arrigoni-99,Arrigoni-00} and functional renormalization\cite{Ledowski-Kopietz-07} support the idea that the warped Fermi-surface is unstable for sufficiently small $t_\perp$, in principle therefore leading to a Fermi-surface modifying transition at some finite value of inter-chain hopping.  While the issue of the Luttinger liquid to Fermi liquid crossover/transition is not yet fully resolved, it is possible to ask a much simpler question: what happens if an array of one-dimensional {\em Fermi liquids} are coupled together? A rather specific example in this direction was the study of coupled edge states (chiral one-dimensional Fermi liquids) in superlattices which exhibit integer quantum Hall effect.\cite{Betouras-Chalker}

While such a toy model may not be realistic for any real materials, advances in laser trapping and cooling technology have led to the rapidly expanding field of trapped ultra-cold atoms, which in the context of condensed matter physics can be thought of as a sort of quantum analogue simulation of a bulk system,\cite{Campo-et-al-07,Ho-Zhou-2010} with unprecedented control over disorder and interactions.  It was suggested a few years ago\cite{2005-Baranov-et-al,Buchler-et-al-07,Micheli-et-al-07} that exploiting the dipole interaction between cold polar molecules or highly dipolar atoms allows further control over effective interactions, in order to build exotic strongly correlated phases. Of particular interest in the present context is the case when the atoms or molecules are fermions.\cite{2005-Baranov-et-al} On the experimental front, there has been much recent experimental progress towards this goal using highly polar $^{40}$K$^{87}$Rb molecules\cite{Ni-et-al-2008} and a fermionic isotope of the highly magnetic atom $^{163}$Dy.\cite{Lu-Youn-Lev-2010} On the theoretical front there has been a flurry of activity.\cite{2008-Miyakawa-Sogo-Pu,Quintanilla-Carr-Betouras-2009,Carr-Quintanilla-Betouras-2009,Fregoso-et-al-2009,2009-Cooper-Shlyapnikov,2009-Huang-Wang,Zhao-et-al-2009,Lin-et-al-2009}

Unlike the long-wavelength scattering induced by Feshbach resonances, dipolar interactions have a power-law dependence on the distance between the interacting particles and a non-trivial dependence on the relative position of the two particles and orientation of the magnetic dipoles. In the presence of a strong polarising field the latter translates into a strongly anisotropic interaction which leads to a spontaneous (though not symmetry-breaking) deformation of the Fermi 
surface.\cite{2008-Miyakawa-Sogo-Pu} Indeed depending on the strength of the dipolar interaction additional, symmetry-breaking (Pomeranchuk) Fermi surface deformations may also occur.\cite{Fregoso-et-al-2009}  Even more interestingly, 
such polarised dipolar gases can in theory be combined with optical lattices to generate non-trivial tailor-made effective Hamiltonians.

In this connection the present authors showed\cite{Quintanilla-Carr-Betouras-2009} that a quasi-one dimensional (quasi-1D) optical lattice could be used to create a system whose phase diagram features Fermi liquid, stripe and checkerboard ground states, as well as a meta-nematic quantum phase 
transition into a state with distorted Fermi surface. The model of Ref.~\onlinecite{Quintanilla-Carr-Betouras-2009} features chains within which there are no interactions, achieved by the alignment of the polarising field at the `magic angle' to the tube direction. For this particular orientation of the field, the interactions 
between particles on different chains are purely repulsive. An experimental realisation of this model would thus furnish an example of the coupled one-dimensional Fermi liquids mentioned above. Interestingly, the spontaneous Fermi surface distortion encountered in this model corresponds also to an interaction-induced change of dimensionality, from quasi-1D (open Fermi surface) to fully 2D (closed Fermi surface) behavior.  An important property of this model is that these various transitions can all happen in different, well separated, regions of phase space.  Thus, each of the phase transitions can be studied independently without its properties being masked by the other ones.

More recently other authors have studied a closely-related model, featuring continuum tubes rather than discrete chains.\cite{2009-Huang-Wang} In this case only the strictly 1D limit was considered but on the other hand the polarising field 
was allowed to point in any direction. This leads to a rich variety of effective interactions and a correspondingly rich phase diagram: in addition to density wave and meta-nematic phases, different superfluid ground states are expected.

The present work extends the theories in Refs.~\onlinecite{Quintanilla-Carr-Betouras-2009,Carr-Quintanilla-Betouras-2009} by considering finite temperature phase transitions in addition to the quantum phase transitions discussed to date.  We also develop a fully analytic theory of the meta-nematic transition in the neighborhood of its critical end point.

The Hamiltonian of the model that we analyze in detail consists of spinless fermions hopping along parallel chains (labelled by $n$), with a weak hopping and an interaction between nearest neighbor chains:
\begin{multline}
\ham = \sum_{i,n} \left\{ -t_\| \left(c^\dagger_{i,n} c_{i+1,n} + c^\dagger_{i+1,n}  c_{i,n} \right) \right. \\
- \left. t_\perp \left(c^\dagger_{i,n} c_{i,n+1} + c^\dagger_{i,n+1}  c_{i,n} \right) 
+  V \rho_{i,n} \rho_{i,n+1} \right\}\label{eq:model}
\end{multline}
where the density
\be
\rho_{i,n} = c^\dagger_{i,n} c_{i,n}.
\ee

The plan for the rest of the paper is as follows: In Section II we will study the meta-nematic transition in full detail, explaining its origins and its link to the Lifshitz transition.   In Section III we will look at the other competing instability of the model to a crystalline phase. The complete physical picture is presented in Section IV.  We finish with a discussion of both the specific properties of this model, and how these may be generalized to a more realistic microscopic model in condensed matter systems.


\section{The meta-nematic transition}

We first consider a phase transition closely related to the Lifshitz transition.\cite{Lifshitz-1960}

\subsection{Self-Energy in self-consistent Hartree-Fock approximation}

We study how the interactions renormalize the spectrum, within the diagrammatic perturbation theory approach.  To do this, we will introduce the Matsubara Green function
\be
G({\bf k},i\omega_n) = \int _0^{1/T} d\tau e^{i\omega_n \tau} \la\la  T_\tau c_{\bf k} (\tau) c_{\bf k}^\dagger \ra\ra,
\ee
where $T_\tau$ is the (imaginary) time ordering operator, $c_{\bf k}(\tau)$ is the annihilation operator in the Heisenberg representation evolving in imaginary time, and $\la\la\ldots\ra\ra$ is the thermal average. In the above expression and in what follows we have set Boltzmann's constant $k_B=1$ and Planck's constant divided by $2\pi$, $\hbar=1$.
Within the diagrammatic perturbation theory, the Green function can be built up from the non-interacting Green functions
\be
G_0({\bf k},i\omega_n) = \frac{1}{i\omega_n-\xi_0({\bf k}) }.
\ee
Here,
\be
\xi_0({\bf k}) = -2t_\| \cos k_x - 2 t_\perp \cos k_y - \mu
\ee
is the bare dispersion.  The full Green function is then given by a Dyson series to be
\be
 G({\bf k},i\omega_n) = 
\frac{1}{i\omega_n - \xi_0({\bf k}) - \Sigma({\bf k},i\omega_n)}.\label{eq:GHF}
\ee
where $\Sigma({\bf k},i\omega_n)$ is the self-energy.

We will calculate the self-energy to lowest order, which corresponds to the Hartree and the Fock terms:
\begin{center}
\includegraphics[width=3in]{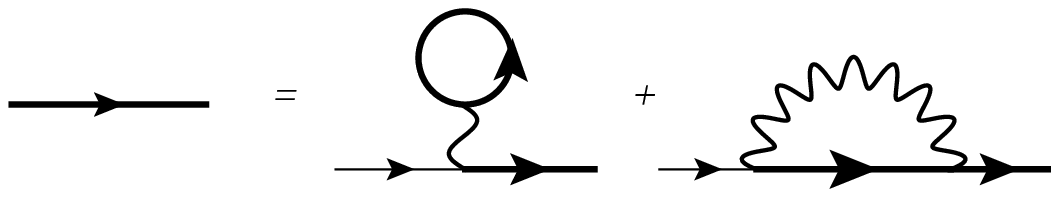}
\end{center}

The first diagram is the Hartree term. It has no dependence on $k_x,k_y$ or $i\omega_n$ and can be trivially evaluated to give
\be
\Sigma_H = -VN/\Omega
\ee
where $N$ is the numer of particles and $\Omega$ is the volume of the system.  This term simply renomalizes the Fermi-energy and will be ignored for now.

The second (Fock) term has more structure.  The diagram corresponds to the expression
\be
\Sigma_F ({\bf k},\omega) = T\sum_{i\epsilon_n} \int \frac{d{\bf q}}{(2\pi)^2}
G_0 ({\bf q} ,i\epsilon_n) V({\bf k}-{\bf q})
\ee
where we have introduced the Fourier transform of the interaction
\be
V\left({\bf p}\right) = 2V\cos\left(p_y\right).
\ee
Now, the Matsubara summation can be performed to give
\be
T\sum_{i\epsilon_n} G_0 ({\bf q} ,i\epsilon_n) = \frac{1}{1+e^{\xi_0({\bf q})/T}}= n({\bf q})
\ee
where $n$ is the Fermi function (with respect to the non-interacting part of the Hamiltonian). We see first that $\Sigma_F$ is independent of energy, and is given by
\be
\Sigma_F (k_x,k_y) = 2V \int \frac{d{\bf q}}{(2\pi)^2}  n (q_x,q_y) \cos (k_y-q_y).
\ee
Note that  $\Sigma_F(k_x,k_y)=\Sigma_F(k_y)$ so the
Hartree-Fock self-energy does not depend on $k_x$. This is a direct
consequence of the interaction potential being $k_x$-independent:
$V(k_x,k_y)=V(k_y)$ and is a specific feature of our particular model.
In the absence of any term breaking the symmetry $k_y\rightarrow -k_y$ (which would give rise to spontaneous currents) we have $n(q_x,q_y)=n(q_x,-q_y)$, so
\be
\Sigma_F (k_y) = 2V \cos k_y \; \int \frac{dq_x}{2\pi} \frac{dq_y}{2\pi} n (q_x,q_y) \cos(q_y).
\ee
Now, from Eq.~\ref{eq:GHF}, we see that the role of an energy independent self-energy is simply to renormalize the bare energy spectrum:
\be
\xi({\bf k}) = \xi_0({\bf k}) + \Sigma({\bf k}).
\ee
Specifically, the effect of this diagram is to renormalize the transverse hopping to
\be
t_\perp^* = t_\perp + V \int \frac{d{\bf q}}{(2\pi)^2} n (q_x,q_y) \cos(q_y).\label{eq:tstar1}
\ee
To make this Hartree-Fock theory self-consistent, we need to take $n(q_x,q_y)$ to be the Fermi function with the renormalized spectrum
\be
n({\bf q}) =\frac{1}{1+e^{\xi({\bf q})/T}}
\ee
where
\be
\xi({\bf q})  = -2t_\| \cos q_x - 2t_\perp^* \cos q_y - \mu.\label{eq:renspec}
\ee
For a variational derivation of these self-consistency equations see Ref.~\onlinecite{Quintanilla-Carr-Betouras-2009}.

While the self-consistent expression for the renormalized transverse hopping, Eq.~\ref{eq:tstar1} looks innocent enough, its solution has a number of surprising features for the anisotropic tight binding model spectrum in question.  Our strategy for the rest of this section proceeds in two steps.  First we will analyze the solutions of Eq.~\ref{eq:tstar1} at zero temperature, where we will discover that far from evolving smoothly, the renormalized hopping undergoes a first order jump at some critical $V$.  We refer to this Fermi-surface shape changing transition as the meta-nematic transition, and we will examine its properties showing that it is the finite interaction version of the well known Lifshitz transition, which as we shall see is its quantum critical end point. 

It is in order here to clarify that the term meta-nematic transition is used in direct analogy with the term meta-magnetic transition, extensively used in recent years.\cite{Perry-et-al-2001,Grigera-et-al-2001} In the first case the transition is caused by the external tuning of the bare tunneling strength whereas in the second case the transition (or crossover) is caused by the tuning of an external magnetic field. Thus in both cases we are dealing with the non symmetry-breaking version of a symmetry-breaking instability (the Pomeranchuk instability in one case and ferromagnetism in the other) in the form of a first-order transition tuned by a symmetry-breaking field (namely, lattice anisotropy and an externally-applied magnetic field, respectively).

In the second step, we will proceed to finite temperature, 
where we will see that this line of first order transitions ends in a critical end-point at a certain $T_c$.    We will also show that this transition is a density of states related effect, and show why we expect it to be robust against higher orders in perturbation theory.

\subsection{The meta-nematic transition at zero temperature}

\begin{figure}
\begin{center}
\psfrag{tperpstar}{$t_\perp^*/t_{\|}$}
\psfrag{V}{$V/t_{\|}$}
\includegraphics[width=3in]{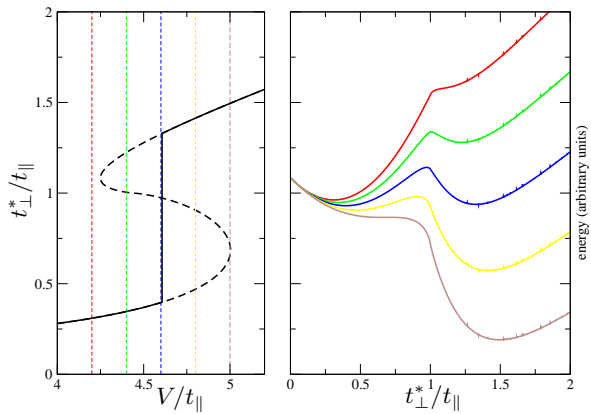}
\end{center}
\caption{[Color online] Left panel: The dashed curve shows the saddle point of Eq.~\ref{eq:tstar1}, for a typical value of $\mu=0$ and $t_\perp/t_{\|}=0.1$.  There is an unstable region where there are three solutions of this equation.    To determine the true ground state solution, we take cuts through this and look at the energy (right panel) for values of $V/t_{\|}=4.2,4.4,4.6,4.8,5.0$ top to bottom.  The local minima correspond to the saddle point solution, while the true minima  allow us to draw on the left the true renormalization of $t^*$, (solid line in first panel) which has a first order jump at some critical $V$. }\label{fig:energyscan}
\end{figure}

In Fig.~\ref{fig:energyscan} we plot a numerical solution of Eq.~\ref{eq:tstar1}.
It is clearly seen that there is a first order jump in the renormalized hopping $t^*$ at some value of $V$.
In order to find the location of this jump, the meta-nematic transition, we must find which solution is stable in the regime where there are  three possible solutions.  This can be done by computing and analyzing  the total free energy of the system.

The non-interacting and interacting parts of the free energy are
\bea
&& \la F_0 \ra = \int \frac{d^2{\bf k}}{(2\pi)^2} \xi_0 ({\bf k}) n({\bf k}) \\
&&-T \int \frac{d^2{\bf k}}{(2\pi)^2} \left\{  n({\bf k}) \ln n({\bf k}) + (1-n({\bf k})) \ln (1-n({\bf k}))\right\}, \nn
&& \la F_1 \ra = \frac{1}{2}\int \frac{d^2{\bf k}}{(2\pi)^2} \int \frac{d^2{\bf q}}{(2\pi)^2} V({\bf q}) n({\bf k}) \left[ 1 - n({\bf k}+{\bf q})\right], \nonumber 
\eea
with the total free energy obtained by adding them together 
\be
\la F \ra = \la F_0 \ra + \la F_1 \ra.
\label{eq:energies}
\ee 
It is worthwhile noting that the equation for the renormalization of the transverse hopping Eq.~\ref{eq:tstar1} is exactly the solution of $\frac{\p F}{\p t_\perp^*} = 0$.  We defer a discussion of higher order terms in the interaction to section V.

In the right panel of Fig.~\ref{fig:energyscan} we plot how the total energy changes with $t^*$ when every other parameter remains fixed.  For $V$ below the transition,  there is a single energy minimum.  Above the lower transition line, a local minimum at higher $t^*$ appears (the middle line in the 3-valued parts of Figs.~\ref{fig:energyscan} is always a local maximum).  Between the two transition lines, this local minimum becomes lower than the previous minimum - and there is a true first order jump in the value of $t^*$.  Above the upper critical line, the  local minimum at lower $t^*$ vanishes.

\begin{figure}
\begin{center}
\psfrag{tstar}{$t_\perp^*/t_{\|}$}
\psfrag{intV}{$V/t_{\|}$}
\psfrag{deltat}{$\Delta t_\perp^*/t_{\|}$}
\includegraphics[width=3in,clip=true]{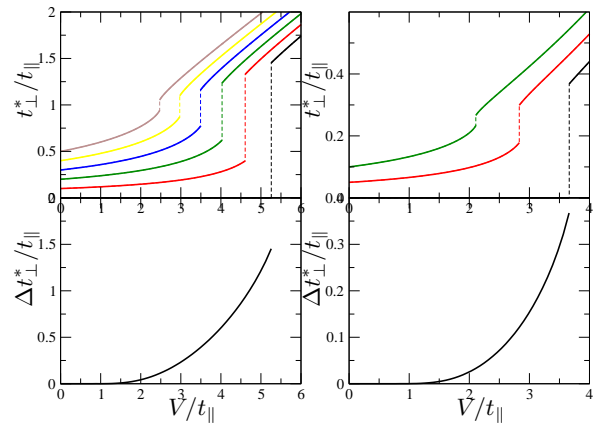}
\end{center}
\caption{[Color online] The renormalized hopping as a function of the interaction strength for $\mu=0.0$ (left) and $\mu=-1.5 t_{\|}$ (right), and various values of the bare interchain hopping $t_\perp$, which can be read off the value of $t_\perp^*$ when $V=0$. The lower panels also show the size of the jump $\Delta t_\perp^*$ as a function of the position $V$ at which it takes place. }\label{fig:tstarVjump}
\end{figure}

We are now in a position to calculate numerically the energy at its minima and find the lowest one.  This allows us to determine and show in Fig.~\ref{fig:tstarVjump}  where the actual transition takes place. In addition, we plot the size of the jump. Note that the jump diminishes as  the value of $V$ where it occurs tends to 0, with a quantum critical end-point at $V=0$.

\begin{figure}
\begin{center}
\psfrag{tstar}{$t_\perp^*/t_{\|}$}
\psfrag{mu}{$\mu/t_\|$}
\includegraphics[width=3in,clip=true]{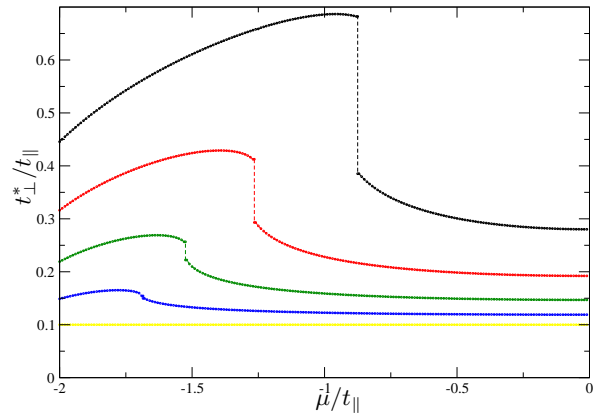}
\end{center}
\caption{[Color online]  The variation of the renormalized hopping $t^*$ with $\mu$ at a fixed bare hopping $t_\perp^{(0)}=0.1t_{\|}$.  The different lines are for different interaction strengths, from bottom to top, $V=0,1,2,3,4$ in units of $t_{\|}$.}\label{fig:tstarmu}
\end{figure}

While $V$ is the natural parameter to vary in order to understand the nature of the transition, much insight can be gained by plotting the renormalized $t^*$ as a function of the chemical potential $\mu$ for fixed bare interchain hopping $t_\perp$ and interaction strength $V$ (in addition, in a cold atom realisation it will be easier to tune $t_{\perp}$ than $V$).  This is done in Fig.~\ref{fig:tstarmu}. Note the very pronounced dependence of $t_{\perp}^*$ on $\mu$ near the transition, even for $V=t_{\|}$, in spite of the first-order jump itself being very small for such small value of the interaction strength.

\begin{figure}
\begin{center}
\psfrag{tperp}{$t_\perp/t_{\|}$}
\psfrag{tjump}{$\Delta t_\perp^*/t_{\|}$}
\psfrag{Vcrit}{$V_c/t_{\|}$}
\includegraphics[width=3in,clip=true]{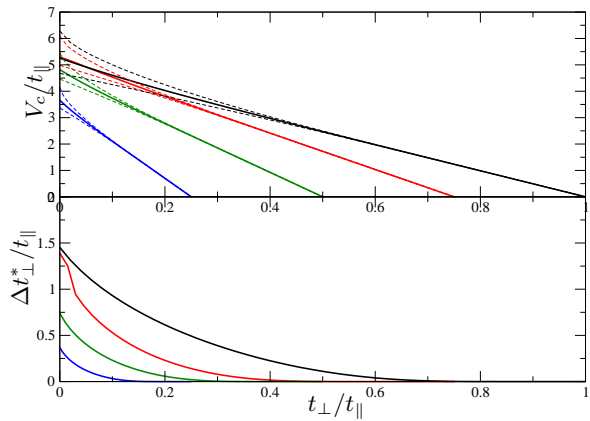}
\end{center}
\caption{[Color online] The upper panel shows the critical interaction strength $V_c$ as a function of the bare interchain hopping $t_\perp$ for four different values of the chemical potential (from top to bottom $\mu/t_{\|}=0.0,-0.5,-1.0,-1.5$).  The dashed lines  in each of these cases bound the region where the energy landscape has more than one local minimum.  The lower panel shows the size of the jump $\Delta t_\perp^*$ as the lines in the upper panel are crossed.}\label{fig:transition2}
\end{figure}

We can now plot the actual zero-temperature phase diagram, showing the single line of first order phase transitions.  This is shown in Fig.~\ref{fig:transition2}. The 2 and 1/2-order Lifshitz transition\cite{Lifshitz-1960} can be regarded as the quantum critical end-point of our meta-nematic first order transition, which we will discuss in more detail in the next subsection.

\subsection{A theory of the quantum phase transition}

So far, all of the results have been obtained by numerical evaluation of the integrals.  We now go one step further, and look at an analytic solution around the quantum critical end point (the Lifshitz transition), where the critical $V_c\rightarrow 0$.  This occurs at the point where the (bare) Fermi surface just touches the edge of the Broullin zone, $t_{\perp}= t_\perp^{QPT}=t_{\|}-\mu/2$.  We write $x=(t_\perp^*-t_\perp^{QPT})/t_\|$ and $x_0=(t_{\perp}-t_\perp^{QPT})/t_\|$, noting that in this notation, $x_0$ will always be negative.  By a direct expansion for small $x$ of the energy, Eq.~\ref{eq:energies}, at zero temperature, we find (we have relegated the details of the derivation to the appendix.)
\be
E\propto x^2\left[ \ln |x| - \frac{1}{2}\right] - 2x\left[ x-x_0-aV\right]\ln|x| - Vb\,x^2\ln^2|x|,\label{eq:GL}
\ee
where $a$ and $b$ are (positive) constants, depending on the chemical potential $\mu$, and we have ignored any terms independent of $x$.  The logarithms reflect the presence of logarithmic singularities in the density of states (van-Hove singularities) at the Fermi level when $x=0$.\cite{Carr-Quintanilla-Betouras-2009}  We note that a similar expansion emphasizing the non-analyticities around the quantum critical end point has been made in a related two dimensional model,\cite{Yamaji-Misawa-Imada-2006} where interactions also drive the Lifshitz transition first order.
The Energy function, Eq.~\ref{eq:GL}, can now be minimized with respect to $x$:
\be
\frac{\p E}{\p x} \propto 2\left[\ln|x|+1\right]\left[-(x-x_0)+V\left(a-bx\,\ln|x|\right)\right].\label{eq:dGL}
\ee
The turning points $\frac{\p E}{\p x}=0$ at $x=\pm 1/e$ (when the first bracket in Eq.~\ref{eq:dGL} becomes zero) are unphysical and an artifact of our small $x$ expansion Eq.~\ref{eq:GL}.  The remaining local minima of \ref{eq:GL} occur when
\be
x-x_0 = V\left[a-bx\,\ln|x|\right].\label{eq:condition}
\ee
It is worthwhile noting that when $V=0$, the only solution of this equation is $x=x_0$, as it must be, and in fact it is this condition that strongly constrained the form of the energy functional, Eq.~\ref{eq:GL}.  Now, at finite $V$, we know that the meta-nematic transition occurs when the energy of the two local minima becomes equal, so that the value of $x$ which minimizes $E$ jumps from one minima to the other.  This occurs at the point $x_0+aV=0$, or
\be
V_c = -x_0/a
\ee
where the energy functional Eq.~\ref{eq:GL} becomes an even function of $x$.  This linear relation between the critical $V$ and the distance from the QPT agrees prefectly with the numerical results in Fig.~\ref{fig:transition2}.  Now, substituting the value of $V_c$ into Eq.~\ref{eq:condition} gives us the solution $x=\pm \exp\left\{ -1/bV_c \right\}$, or
\be
\Delta t^*_\perp =2 e^{-\frac{1}{bV_c} }.
\ee
This non-perturbative expression for the jump in the renormalized transverse hopping, which acts as an effective `order parameter', again agrees well with the numerics in Figs.~\ref{fig:tstarVjump} and \ref{fig:transition2}. We note that the expression is of the same form as the well-known one giving the finite binding energy of a Cooper pair in the presence of and arbitrarily weak attractive interaction. Its meaning is that the 2 and 1/2-order Lifshitz transition turns first order under the effect of interactions in a non-analytic way, making it an idealisation never truly realised in any real (and therefore, interacting) system. Note that recent experimental evidence in favour of a Lifshitz phase transition in Na$_x$CoO$_2$ seems to suggest that the transition is indeed of first order in this material.\cite{Okamoto-Nishio-Hiroi-2010} Other interaction-induced non symmetry-breaking phase transitions involving a change of topology of the Fermi surface have been discussed in Refs.~\onlinecite{Edwards-2006,Quintanilla-Schofield-2006,Yamaji-Misawa-Imada-2006,Yamaji-Misawa-Imada-2007}.


\subsection{Finite temperature}

\begin{figure}
\begin{center}
\psfrag{Vint}{$V/t_{\|}$}
\psfrag{tstar}{$t_\perp^*/t_{\|}$}
\includegraphics[width=3in,clip=true]{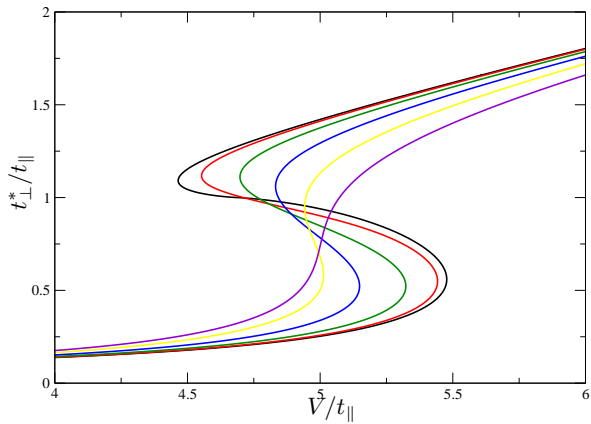}
\end{center}
\caption{[Color online]  The solution of the saddle point Eq.~\ref{eq:tstar1}, at finite temperature (from outside to inside $T/t_{\|}=0.0,0.1,0.2,0.3,0.4,0.5$) for a typical value of $\mu=0$ and $t_\perp/t_{\|}=0.05$.   As temperature is raised, the size of the region with more than one local minima decreases, and eventually disappears altogether when we reach $T_c$, the second-order critical end-point of the meta-nematic transition. }\label{fig:tstarVT}
\end{figure}

In this section we study the effects of temperature on the meta-nematic transition.  The equation for the renormalization of $t_\perp$ is the same as for the zero temperature case, but with the zero temperature distribution replaced by the finite temperature Fermi distribution.
In this case we perform the integrations numerically.  A representative  cross section of the results are plotted in Fig.~\ref{fig:tstarVT}.  As expected, by increasing temperature, the size of the first order jump gets smaller,
until it finally vanishes at a 2nd order critical point at some $(T_c,V_c)$.

\begin{figure}
\begin{center}
\psfrag{Vint}{$V/t_{\|}$}
\psfrag{Temp}{$T/t_{\|}$}
\psfrag{deltatstar}{$\Delta t^*_\perp/t_{\|}$}
\includegraphics[width=3in,clip=true]{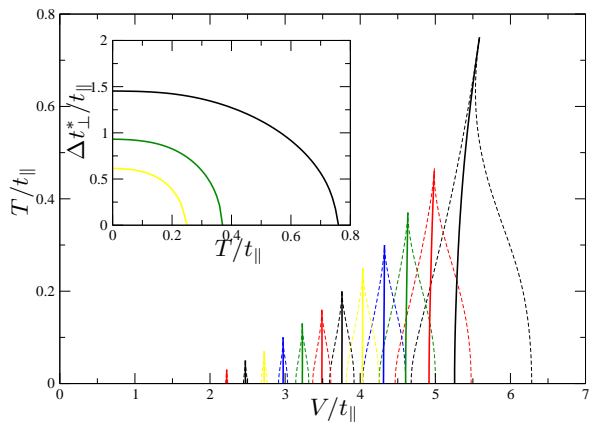} 
\end{center}
\caption{[Color online] The critical line in the $V-T$ plane, for an example value of chemical potential $\mu=0$.  From right to left, the bare hoppings are $t_\perp/t_{\|} = 0.0,0.05,0.1,0.15,0.2,0.25,0.3,0.35,0.4,0.45,0.5,0.55$.  The dashed lines enclose the region of interaction strengths $V$ for which the free energy has more than one local minimum.  Inset: The size of the jump $\Delta t_\perp^*/t_{\|}$ as we cross the critical line, as a function of temperature $T$.  From outside to inside, the values of $t_\perp/t_{\|}=0.0,0.1,0.2$.  }\label{fig:PhaseVT}
\end{figure}

We find the position of the jump within this unstable region by free energy considerations identical to the zero-temperature case.  This is plotted in Fig.~\ref{fig:PhaseVT} as a function of temperature $T$ for different values of $\mu$ and the bare hopping $t_\perp$.
We note that unlike the quantum case, the classical second-order phase transition (where the jump vanishes at finite $T$) has a regular free energy expansion as the van-Hove logarithms are smoothed over by temperature - leading to the standard analytic mean-field behavior whith the expected threshold temperature dependence:
\be
\Delta t^*_\perp \propto (T_c-T)^{1/2},
\ee
which is shown as the inset to Fig.~\ref{fig:PhaseVT}.

\section{The density wave transition}

The model, Eq.~\ref{eq:model} also has an instability to a density wave (DW) state, which has broken translational symmetry characterized by a non-uniform density:
\be
\rho_{DW}(x,y) = \rho_0 + A \cos(2k_F x + \pi y).
\ee
 This arises from the quasi-one dimensional nature of the model, which has an almost nested Fermi surface (which becomes perfectly nested in the limit $t_\perp\rightarrow 0$ or $\mu=0$).  While the physics of this transition is well established, it is important to investigate it in more detail in this case, in order to determine the regions in parameter space where the new meta-nematic transition is not eclipsed by this DW transition.

We will estimate the transition line between the DW state and the `normal' phase via the Random Phase Approximation (RPA). We note that this apporach is valid only for determining where the DW instability takes place, and not to describe the ordered state beyond the instability line. For that, one could add a translational symmetry breaking mean field and redo the Hartree-Fock theory. This would have the advantage of furnishing a description of the symmetry broken phase as well. However, the presence of a DW incommensurate with the lattice creates many calculational difficulties, and the instability condition obtained in this way is equivalent to that obtained within the RPA. Furthermore, the use of RPA emphasizes the two-particle nature of the DW instability, in contrast to the one-particle nature of the MN transition; thus each are controlled by different properties of the Fermi surface (see section V).

We start with the density-density correlation functions in the Matsubara time representation:
\be
X({\bf k},\tau) = i \la\la  T_\tau \rho({\bf k},\tau) \rho^\dagger ({\bf k},0)\ra\ra.
\ee

In the non-interacting case, this is given by the particle-hole bubble:
\begin{eqnarray}
&&X^0 ( {\bf k}, \omega) \nn
&=& -\int \frac{d{\bf q}}{(2\pi)^2}
\int \frac{d\epsilon}{2\pi i}
G^0 ( {\bf k}+{\bf q}, \epsilon+\omega) G^0 ({\bf q},\epsilon).
\end{eqnarray}
Treating the interactions via the RPA then gives us:
\begin{equation}
X ({\bf k},\omega)
= \frac{X^0({\bf k},\omega)}{1 + V(k_y) X^0 ({\bf k},\omega)}.
\label{rpaeq}
\end{equation}

The momentum dependent interaction $V(k_y)$ is given by
\be
V(k_y) = 2V \cos k_y.
\ee
In particular, setting $k_y=\pm \pi$, this is negative, and could lead to an instability.  In fact, the Stoner criterion is exactly looking for poles in response functions at zero frequency, $\omega=0$.  Looking at Eq. \ref{rpaeq}, we see that this happens when
\be
2 V X^0 ({\bf k}, \omega=0) = 1.
\ee
The imaginary part of  $X_0$ vanishes at $\omega=0$.
Taking the real part gives the Lindhard formula for density-density response:
\be
\Re X_0 ({\bf k},\omega) = \int \frac{d{\bf q}}{(2\pi)^2} \frac{n({\bf q}) - n({\bf k}+{\bf q})}{\omega - \xi({\bf k}+{\bf q}) + \xi ({\bf q})},
\ee
where $n({\bf k})$ is the Fermi-function.  We can include the effects of the Hartree-Fock self-energy in this expression by taking $\xi({\bf k})$ to be the renormalized energy spectrum, given by Eqs. \ref{eq:tstar1} and \ref{eq:renspec}, rather than the bare tight-binding spectrum.  The most important point is that so long as the renormalized inter-chain hopping remains small, $X_0({\bf k},0)$ is only very weakly dependent on $k_y$.

In general, the Lindhard function (and hence the Stoner criterion for the phase boundary) must be evaluated numerically, however there are some limits where we can gain insight analytically.  Let us consider first the case where the renormalized interchain hopping $t_y^*$ remains zero.  Therefore, near the Fermi surface, the nesting condition $\xi({\bf k}+{\bf Q})\approx-\xi({\bf k})$ is satisfied for the nesting vector ${\bf Q}=(2k_F,\pi)$.  This gives
\be
X_0({\bf Q},\omega=0) \sim \rho(\epsilon_F) \ln \left( \frac{\epsilon_F}{T} \right),
\ee
where $\rho(\epsilon_F)$ is the density of states at the Fermi surface, and $\epsilon_F\approx 2t^\|+\mu$ is the chemical potential as measured from the bottom of the band.
Thus the Lindhard function has a logarithmic divergence at wavevector ${\bf Q}$ at zero frequency and zero temperature.  This means that the ground state is indeed a density wave for any finite interaction strength, with a critical temperature estimated from the Stoner criterion:
\be
T_c \sim \epsilon_F e^{-1/\rho(\epsilon_F)V}.
\ee
This means that with zero interchain hopping, the metanematic transition of the previous section will never be seen, as it will always be preempted by an ordering transition.

For finite (but small) interchain hopping, the perfect nesting is lifted (so long as we stay away from the particle-hole symmetric point $\mu=0$) and there is no longer a singularity:
\be
X_0({\bf Q},\omega=0,T=0) \sim \rho(\epsilon_F) \ln \left( \frac{\epsilon_F}{t_\perp^*} \right),
\ee
meaning that a finite interaction
\be
V_c \sim \frac{1}{\rho(\epsilon_F) \ln \left( \frac{\epsilon_F}{t_\perp^*} \right)}
\ee
must be obtained before the instability to the density-wave state occurs.  This gives us a region of parameter space where as the interaction is turned up, the metanematic jump takes place before the DW transition, so the metanematic transition can be seen.

To go beyond these simple limits requires a numerical evaluation of the Lindhard function.  In the next section, we will plot the phase boundary of the DW transition alongside the phase boundary of the metanematic transition allowing the region in phase space where the metanematic transition `wins' to be clearly seen.  Before going on to this however, we will take a look at corrections to the RPA, which  in our particular model will be proven to be small.  We will then look at a strong coupling expansion which will complement the weak coupling perturbative analysis in this section, before finally giving a completely different derivation of the DW instability for the special case of $t_\perp=0$.


\subsection{Vertex Corrections}

In this subsection we go beyond the RPA and study the vertex corrections (note the fact that RPA becomes exact for a strictly 1D linear spectrum\cite{Dzyaloshinskii-Larkin-1974}).  First we examine the case where the renormalized hopping remains zero, so the Green functions are independent of transverse momentum, and the interaction lines are independent of the longitudinal momentum.

\begin{figure}
\begin{center}
\includegraphics[width=3in]{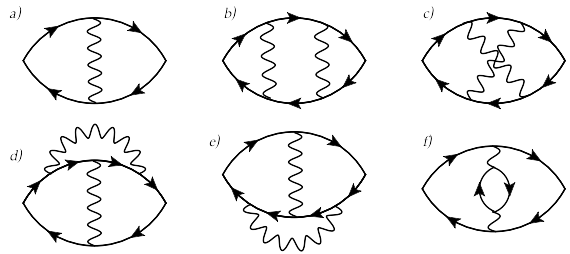}
\end{center}
\caption{The first and second order vertex corrections}\label{fig:vertex}
\end{figure}

The first order correction is shown in Fig.~\ref{fig:vertex} (a).  We note that the particle lines are dressed by the Hartree-Fock self-energy, see Eq.~7. This takes care of all
first-order vertex corrections except for that represented by the diagram $(a)$, which is given by 
\bea
&& X^1(k_x,k_y,\omega) = 
\int \frac{d{\bf q}}{(2\pi)^2} \int \frac{d\epsilon_1}{2\pi} \int \frac{d{\bf p}}{(2\pi)^2} \int \frac{d\epsilon_2}{2\pi} \nn 
&& \times G^0(k_x+q_x,\epsilon_1+\omega) 
G^0(k_x+q_x+p_x,\epsilon_1+\epsilon_2+\omega) 
 \nn && \times G^0(q_x,\epsilon_1) G^0(q_x+p_x,\epsilon_1+\epsilon_2) V(p_y) = 0.
\eea
The result is zero due to the integration over transverse momentum, $\int dp_y V(p_y)=0$.  We emphasize that this is a special feature of the model under consideration. 

The second order diagrams are also shown in Fig.~\ref{fig:vertex}.  Diagrams $(b)$--$(e)$ will all be zero for the same reason as the first order diagram.  Diagram $(f)$ is therefore the first correction to the RPA
\bea
&& X^2(k_x,k_y,\omega) 
= \int \frac{d{\bf q}}{(2\pi)^2} \int \frac{d\epsilon_1}{2\pi} \int \frac{d{\bf p}}{(2\pi)^2} \int \frac{d\epsilon_2}{2\pi} \times \nn
&& G^0(k_x+q_x,\epsilon_1+\omega)  G^0(k_x+q_x+p_x,\epsilon_1+\epsilon_2+\omega) \nn
&&G^0(q_x,\epsilon_1)  G^0(q_x+p_x,\epsilon_1+\epsilon_2) X^0 (p_y,\epsilon_2 )[V(p_y)]^2.
\eea

While this is non-zero, note firstly that the entire expression is independent of $k_y$, it is smaller than $X^0$ by a factor $(V/t_\|)^2$ and most importantly, does not have a divergence at $k_x=2k_F$ (basically the divergence is cured by the extra integrations).  In the region of interest $k_x \approx 2k_F, \omega=0$, it can therefore be considered as a small perturbation to $X^0$ within the RPA and thus does not have to be taken into account to get the essential physics correct.

All of this is strictly true for $t_\perp=0$. Nevertheless, it is easy to check that when $t_\perp\ne 0$, these vertex corrections form a power series in $t_\perp/t_\|$. Therefore as long as the renormalized value of $t_\perp$ remains small, the RPA is a very good approximation.
Similar features has been seen e.g.  in a model of stacked edge states in the case of the integer quantum Hall effect.\cite{Betouras-Chalker}


\subsection{Strong coupling limit}

While the smallness of the vertex corrections means that RPA is formally a very good approximation for the model in question, it still remains a weak coupling perturbation expansion.  We can complement these results by showing that in the strong coupling limit, 
the ground state is indeed a $(2k_F,\pi)$ density wave.  This demonstrates the qualitative correctness of the phase diagram we will present in the next section, even if the RPA does not 
estimate with high precision the position of the transition line for large interaction strengths. 

We can proceed by first taking the kinetic part of the Hamiltonian Eq.~\ref{eq:model}  to zero $t_\| = t_\perp=0$. In this limit, we can rewrite the Hamiltonian as:

\bea
\ham_0 &=& V \sum_{i,n} \rho_{i,n} \rho_{i, n+1} \\
&=& \frac{1}{2}V \sum_{i,n} \left[ (\rho_{i,n}+ \rho_{i, n+1}) - (\rho_{i,n}- \rho_{i, n+1})^2 \right]  \nonumber
\eea

\noindent where we have used the property $\rho_{i,n}^2 =\rho_{i,n}$. From the above form of the Hamiltonian it becomes obvious that a uniform distribution of  fermions over the chains is not energetically favorable.
To illustrate this, consider the half-filled case, where half of the sites are occupied in the system with M chains each one having $N_{c}$ sites, therefore total fermions $N=M N_c/2$.
If the real space pattern is such that every column (perpendicular to chains)  has every second site occupied then the total energy is zero.
 Similarly this result stays the same if we choose any particular way how to ``combine", or ``lock"
these columns and form the chains. Therefore there is a macroscopic degeneracy of $2^{N_c}$.

It is interesting to map the problem onto an Ising Hamiltonian by allowing the values of a pseudospin S to be +1 in the presence of a fermion at a given lattice site and -1 in the absence of it.
Then we can transform the Hamiltonian to the two-dimensional antiferromagnetic Ising with coupling constant only in one direction. This can be done by the transformations:
$S_{i,n}=2 \rho_{i,n}-1$ and $\rho_{i,n}=1/2 (S_{i,n}+1)$. The Hamiltonian then maps onto:
\be
H=\frac{V}{4} \sum_{i,n} S_{i,n} S_{i,n+1}
\ee
which has an antiferromagnetic ground state in rows vertical to chains and therefore the same degeneracy as shown above.

If we turn on the kinetic energy part we observe that to first order in perturbation theory, we cannot have non-zero matrix elements among the degenerate ground state manifold; 
then second order processes of order $\frac{t^2}{V}$ must be taken into account. Therefore the degeneracy is lifted by any infinitesimal amount of kinetic energy (tunneling strength) which, to second order in perturbation theory,
selects the checkerboard configuration. It is a version of the ``order by disorder" phenomenon;  order is developed by increasing a quantity associated to disorder (kinetic energy). To second order in the tunneling strength the energy gain is:

\bea
\ham_k  = &-& \sum_{i,n} (\frac{t_{\|}^2}{2V} c^\dagger_{i,n} c_{i+1,n}c^\dagger_{i+1,n} c_{i,n}  \\
&+& \frac{t_{\perp}^2}{V} c^\dagger_{i,n} c_{i,n+1}c^\dagger_{i,n+1} c_{i,n} )
\eea

\noindent and it favors the checkerboard pattern. This is precisely the DW (2k$_F$, $\pi$) that was found using RPA calculations. In the pseudospin language, the Ising Hamiltonian acquires an additional term which takes into account the second order tunneling:
\be
H= \sum_{i,n} [ (\frac{V}{4}+\frac{t_{\perp}^2}{V}) S_{i,n} S_{i,n+1} + (\frac{t_{\|}^2}{2V}) S_{i,n} S_{i+1,n} ].
\ee

 At less than half filling (the physics at greater than half filling will be identical through a particle-hole transformation), the ground state degeneracy in the absence of kinetic energy is even worse as there exist extra unoccupied sites which could occur anywhere within the previous manifold of states. However, now first order processes in hopping are allowed. Including the hopping along the chain to first order, one can think of the ground state as being stacks of particles as delocalized as possible, but avoiding contact with other particles above or below. For example, at $1/4$ filling, each particle will be delocalized over two sites in the chains, these stacks then forming an identical problem to the single-site columns in the half filled case. Of course at incommensurate fillings, the particles will be delocalized over a fractional number of sites - but the ultimate physics will remain the same. As before, finally taking second-order processes into account removes the mass-degeneracy and isolates the `crystalline' phase as the unique ground state. The precise determination of the possible favored ground state configurations beyond this intuitive argument requires more detailed computational work which is beyond the scope of the present work and the illustrated physics.


\subsection{The limit of zero interchain hopping}

In the model Eq.~\ref{eq:model} for the specific case $t_\perp=0$, we can study the model in a completely different way, as the non-interacting part is now strictly one-dimensional.   This allows us to single out the most important components of the interaction and look at their flow under the renormalization group (RG) 
rescaling of the system.\cite{Solyom-1978}

On each of the chains, one can make the chiral decomposition:
\be
c_{i,n} \rightarrow e^{-ik_F x} L_n (x) + e^{ik_F x} R_n (x)
\ee
where $x=ai$, $a$ being the lattice spacing, $n$ indicates the chain number and $L$ and $R$ are slowly varying fields near the left and right Fermi points respectively.  
By linearizing the spectrum around the two Fermi points, the kinetic part of the Hamiltonian becomes
\be
\ham_0 = -i v_F \sum_n \int dx \left( R_n^\dagger (x) \p_x R_n (x) - L_n^\dagger(x) \p_x L_n(x) \right).\label{h1}
\ee
where $v_F$ is the (bare) Fermi velocity, and the interaction term becomes
\be
\ham_I = \sum_{nm} \int dx \left\{ g_{1,nm} R^\dagger_n L _n L^\dagger_m R_m
+ g_{2,nm} R^\dagger_n R_n L^\dagger_m L_m \right\},
\ee
where $g_{2,nm}$ is the {\em forward scattering}, and $g_{1,nm}$ is the {\em back scattering} between particles on chains $n$ and $m$.  In terms of the bare parameters of the Hamiltonian, $g_{1,nm}=g_{2,nm}=V$ for $nm$ nearest neighbors and zero otherwise - however under RG, effective further neighbor scattering terms are generated, so we must retain them in the Hamiltonian.  We note that there is also inter-chain interactions involving only right (or left) moving particles - however these are known not to affect the RG flow\cite{Solyom-1978} so we will not consider them.

If only the forward scattering ($g_2$) terms were present, this model can be exactly solved by the Bosonization technique.\cite{Carr-Quintanilla-Betouras-2009}  There is a transition to a smectic phase\cite{Quintanilla-Carr-Betouras-2009} at large $V\sim v_F$.  As this transition will only be visible in a slightly perturbed version of the model described by Eq.~\ref{eq:model} with longer range interactions\cite{Quintanilla-Carr-Betouras-2009} which we have previously discussed, we will concentrate here on the effect of the $g_1$ backscattering terms.

While such models can still be bosonized in the presence of inter-chain backscattering (see e.g. Ref. \onlinecite{Carr-Tsvelik}), the resulting bosonic Hamiltonian is non-linear and not easy to deal with.  We therefore elect to remain in the fermionic representation, where the one-loop RG equations can be easily derived:
\bea
\frac{d g_{1,nm}}{dl} &=& -2 g_{1,nm}\left( g_{2,nm} - g_{2,nn}\right) - \sum_{i} g_{1,ni}g_{1,in} \nn
\frac{d g_{2,nm}}{dl} &=& -g_{1,nm}^2.\label{RG1}
\eea
where $l=-\ln \Lambda$, with $\Lambda$ being the ultra-violet cut-off of the theory.  These are very similar (although not identical) to the related problem of coupled {\em spin-full} chains.\cite{Gorkov-Dzyaloshinskii,Mihaly-Solyom}

Following Ref.~\onlinecite{Mihaly-Solyom}, we can now see that while the bare couplings may be nearest neighbors only, the RG due to the sum in Eq.~\ref{RG1} generates couplings between all pairs of chains.  Furthermore, due to this sum, the backscattering amplitudes renormalize much quicker than their forward scattering counterparts\cite{Mihaly-Solyom}, so to leading order one can neglect the renormalization of $g_2$, and deal with the far simpler equation:
\be
\frac{d g_{1,nm}}{dl} = - \sum_{i} g_{1,ni}g_{1,in} \label{RG2}
\ee
We note that effects of $g_2$ were dealt with in Ref.~\onlinecite{Menyhard} which concludes that the only interesting effects of $g_2$ come when one has a bare intra-chain backscattering amplitude, something we neither have by construction of our interaction, nor can have due to the spinless nature of the model.  We therefore will deal only with Eq.~\ref{RG2}, which are easily separated by a Fourier transform
\be
g_1(q_\perp) = \frac{1}{N} \sum_m e^{i(n-m)q_\perp} g_{1,nm}
\ee
which gives us
\be
\frac{ \p g_1(q_\perp)}{\p l} = - g_1^2(q_\perp).
\ee
Hence any parts of $g_1(q_\perp)$ which begin negative renormalize to strong coupling.  Now, the bare backscattering amplitude from nearest neighbors is
\be
g_1^{\rm bare}(q_\perp) = 2V\cos(q_\perp),
\ee
so under RG, the component of the interaction which hits strong coupling first (after which the RG must be stopped) is the $q_\perp=\pi$ part of the {\em backscattering} amplitude, or in other words, the $(2k_F,\pi)$ component of the interaction.  This signifies the formation of a $(2k_F,\pi)$ DW\cite{Mihaly-Solyom,Solyom-1973} as we expected.


\section{Full phase diagram and effect of trapping potential}

\subsection{Full phase diagram}

\begin{figure}
\begin{center}
\psfrag{tperp}{$t_\perp/t_{\|}$}
\psfrag{Vint}{$V/t_{\|}$}
\psfrag{Temp}{$T/t_{\|}$}
\includegraphics[width=3in,clip=true]{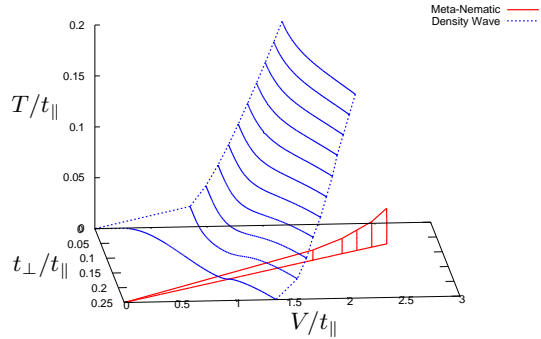} 
\end{center}
\caption{[Color online] Plot of the density-wave and meta-nematic critical surfaces for $\mu/t_\|=-1.5$.  The plots are qualitatively similar for other values of $\mu$.}\label{fig:PhaseDiagramFull}
\end{figure}

We now can draw a complete phase diagram of the model, including both possible instabilities - which is plotted in Figure~\ref{fig:PhaseDiagramFull} for a typical value of $\mu/t_{\|}=-1.5$.  The plots are qualitatively similar for other values of the chemical potential.  From the plot, it may seem that the instability to a crystalline pattern dwarfs the meta-nematic instability, however it is also clear that there will always be a region in phase space where the meta-nematic transition can be observed.  Furthermore, we should add that while in our simple model, the density-wave is usually the dominant instability, the two transitions come from fundamentally different physics: the meta-nematic transition from the van-Hove singularity in the density of states, and the density-wave transition from a near-nested Fermi surface.  Therefore, destroying the nesting by for example adding a next-nearest-neighbor hopping term; or frustrating the interaction will both lead to a suppressed tendency to enter the DW phase whilst barely affecting the meta-nematic one.  However, the best way to enhance the MN phase relative to the DW is open for further investigation.

As noted in the Introduction the separation in phase space between the MN and DW phase transitions is one of the salient features of this particular model, with the phase diagram in Fig.~\ref{fig:PhaseDiagramFull} showing clearly the existence of such regions of phase space where either one or the other instability takes place. On the other hand, in those other regions where the two different instabilities become too close together, our calculational techniques break down, as the mutual competition between each of them must be carefully taken into account. We discuss this point further in Section V below. For the time being we simply must bear in mind that the plotted phase diagram is less reliable around the latter regions.

\subsection{Density profiles in a trapping potential}

\begin{figure}
\begin{center}
\psfrag{tstar}{$t_\perp^*$}
\includegraphics[width=3in,clip=true]{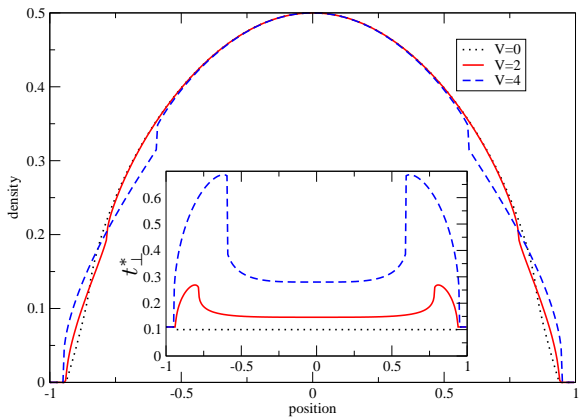}
\end{center}
\caption{[Color online] The local density as a function of position of our model within a harmonic trap (treated within the LDA).  The parameters are adjusted so that the density in the middle of the trap is exactly half filled, and the bare transverse hopping $t_\perp^{(0)}=0.1$.  We note that the variation of the effective chemical potential across the trap makes the meta-nematic transition visible as a jump in the density, with no special tuning of parameters.   Inset: the variation of the 'order parameter' $t_\perp^*$ aginst position in the trap.}\label{fig:trap1}
\end{figure}

Having discussed the properties of transitions in a general context, we now turn specifically to one potential realization of the model: dipolar fermions in an anisotropic optical lattice.\cite{Quintanilla-Carr-Betouras-2009}  In a cold atoms experiment, in addition to the optical lattice,  there is also a harmonic trapping potential:
\be
V(r) = \alpha r^2,
\ee
where $\alpha$ is some constant dependent on the experimental setup.
Within the local density approximation (LDA), this can be theoretically approximated as an inhomogenous chemical potential
\be
\mu _{\text{eff}}(r) = \mu-V(r) = \mu - \alpha r^2.
\ee
From the previous formulas, we can therefore plot the density profile, and profile of renormalized hopping) within the trap, which is shown in Fig.~\ref{fig:trap1}.  We note that in this Figure, we show only 
the metanematic transition and not any possible DW phases: in fact for the rather strong interaction strengths shown in the figure the DW phase will have a dominant presence in the phase diagram. However on reducing $V$, the DW will go away, leaving only the metanematic jump shown.   This jump will be present for any interaction strength $V$, although for small $V$, the size of the jump becomes exponentially small, which is why it is artificially enhanced in the figure, by plotting the profile for large values of $V$ (see also the discussion in the previous section about the relation between the DW and MN transitions).

The interesting feature of these steps in the density profile, is that they signal the separation of two {\em compressible} phases, \cite{Shin-et-al-2008} and do not represent the more common compressible/incompressible boundary.  
The only difference between the phases on either side of the jump is the topology of the Fermi surface: on one side it is closed, while on the other it is open.  This difference is, however, possible to measure in optical lattice experiments by direct imaging of the Fermi surface.\cite{Bloch-2005,Kohl-et-al-2005}

\section{Discussion}

We have examined in detail two different phase transitions associated with quantum liquid crystal phases of matter, which appear in the same physically realizable model.  The first, the meta-nematic transition which is non-symmetry breaking but has strong ties to both the Pomeranchuk and the Lifshitz transitions while the second crystallization transition is a standard symmetry breaking transition.  We have shown that both transitions may be found in a simple anisotropic lattice model of spinless fermions, which may be realized with ultra cold dipolar atoms in an optical lattice.

The Pomeranchuk mechanism of spontaneous deformation of the Fermi surface is a straight-forward generalisation of the Stoner mechanism of itinerant ferromagnetism\cite{Pom58} (very much in the
same way as unconventional pairing generalises s-wave pairing in the BCS theory of superconductors). This begs the question why, unlike itinerant ferromagnetism,
which is ubiquitous, there is so little evidence of Pomeranchuk and other spontaneous deformations of the Fermi surface, such as the meta-nematic transition discussed here, in real systems (in contrast, many unconventional
superconductors exist in addition to those conforming to the original form of BCS theory). One insight is offered by the mean field analysis of central interactions leading to Pomeranchuk
instabilities.\cite{Quintanilla-Schofield-2006} It reveals that unlike the Stoner instability, which may be induced by a contact interaction, its higher-angular momentum generalisations, namely the Pomeranchuk instabilities, require a repulsive interaction potential
$V(r)$ with non-trivial features at a characteristic distance $r_0 >~ r_s$, where $r_s$ is the average distance between particles. Under such conditions other instabilities, involving the breaking of
translation symmetry, are also expected. In particular there is a competition between Fermi surface deformations and more conventional, density-wave instabilities leading to smectic order (stripes) or
crystallisation. A similar situation is found in mean field theories of Fermi surface deformations on the square lattice:\cite{2001-Valenzuela-Vozmediano,2008-Quintanilla-et-al} it is found that repulsion between fermions sitting on nearest-neighbour
sites can lead to Pomeranchuk deformations of the Fermi surface but that this tendency is strongest near half filling where other, real-space instabilities are also enhanced. The question of which type
of order prevails in this case is hard to answer. In general, it will be model-specific and require a quantitative asessment of the role of fluctuations going beyond mean field theories.
It is worth mentioning in passing, that recently, there was a rigorous study of the effects of van Hove points on the properties of interacting fermionic systems, using functional renormalization group (FRG) techniques.\cite{Feldman-Salmhofer} There it was shown
that the logarithmic singularity in the density of states caused by the zero of the bare Fermi velocity gives rise to new marginally relevant terms in the renormalization group equations and leads to singularities in the analytic properties of electronic self-energies.

In the present work we have considered an \emph{anisotropic} square lattice with anisotropic interactions and we have found that in this case the situation is much more clear-cut. As discussed above, the
density wave and meta-nematic instabilities are driven by two separate aspects of the band structure, namely nesting and the van Hove singularity in the density of states, respectively. As is well known,
these two features coincide for a square lattice, but as soon as $t_{\perp} \neq t_{\|}$ they diverge: while perfect nesting still occurs at exactly half filling, $\mu = 0$, the singularity splits into
two separate features at $\mu=\pm 2 |t_{\|}-t_{\perp}|$. As a result of this, for the anisotropic model considered here there is a broad region of the phase diagram where meta-nematic order clearly prevails over the tendency to form a density wave. 

This separation of features also partially justifies the use of the Hartree-Fock approximation - far enough away from the density-wave instability, the susceptibility of the model remains small, so the interaction is only weakly renormalized by higher order terms, leading to only small corrections to the Hartree-Fock results.  As the two different instabilities come closer together in phase space (as would always happen near the Pomeranchuk instability in lattice systems), the interplay between the FS-deforming transition and the crystalline transition must also be considered,  for example using the fluctuation-exchange approximation (FLEX).\cite{Bickers-Scalapino-White-1989}  A closely related competition between spin density wave formation and superconductivity has recently been studied in the context of the pnictide superconductors.\cite{Chubukov-Efremov-Eremin-2008}  Such a study however is beyond the scope of the present work.

We also note that even far from the crystalline transition, the Hartree-Fock approximation neglects all fluctuations, and it is natural to ask what role such fluctuations may play in the meta-nematic transition.  We expect that as this transition is driven by a van-Hove singularity in the density of states, fluctuations which can be described by Fermi-Liquid physics will not qualitatively change the picture.  The reason is that so long as the energy distribution function at $T=0$ has a jump at the Fermi energy, the same logarithmic divergences as we have considered will appear in the perturbation series, which will lead to the same physics, and the effect of the fluctuations will be limited to the renormalization of some parameters.

Interestingly, this separation of features due to anisotropy is somewhat reminiscent of the cuprate high-temperature superconductors, where the existence of stripe order is well-established in tetragonal La$_{1.48}$Nd$_{0.4}$Sr$_{0.12}$CuO$_4$ and La$_{1.875}$Ba$_{0.125}$CuO$_4$ [Ref.~\onlinecite{2005-Abbamonte-et-al}
and references therein] while evidence is accumulating of a nematic state in YBa$_2$Cu$_3$O$_{6+x}.$\cite{2002-Ando-et-al,2008-Hinkov-et-al} The CuO chains present in the latter material constitute a very strong orthorhombic perturbation. One way of interpreting the date is that
the resulting breaking of the $D_4h$ symmetry of the crystal lattice drives away a density wave instability that competes with the nematic order, allowing the latter to prevail in the case of YBCO.

We finally note that in the general case when interactions between next nearest neighbours on different chains become important the density wave instability described here may take place with wave vector ${\bf Q}=(0,\pi)$ (leading to a stripe phase), instead of $(2k_F,\pi)$
(checkerboard). The stripe instability becomes preferable due to a frustration of the nearest neighbour interaction.\cite{Quintanilla-Carr-Betouras-2009} Similarly in some cuprates a checkerboard pattern is observed instead of stripes or
nematic order [Ref.~\onlinecite{2008-Wise-et-al}, and references therein].

In summary, we have studied in detail a toy model which shows a very rich phase diagram and is a playground for many phases which are sought after by the community of correlated fermions.  We have demonstrated within this model the fundamentally different properties of the Fermi surface that lead to crystallization and a Fermi-surface-modifying transition.  We must add that while our analysis shows the presence of a meta-nematic transition, it is exponentially small in the interaction strength, and therefore presents a formidable task at the moment to measure it experimentally. At larger interactions of course, the perturbation theory needs further supplementary work by treating both the meta-nematic transition and the density wave instability on the same footing with e.g. FLEX, or
 with strong coupling techniques such as FRG. We hope that this work will stimulate more studies in this direction, both experimentally and theoretically.

The authors would like to acknowledge useful discussion with P. Kopietz, N. Shannon and A. Lichtenstein.  Much of this work was done while STC was at the University of Birmingham supported by EPSRC grant no. EP/D031109. Part of JJB's work was done at the University of St Andrews where he was supported by a SUPA Advanced Fellowship. JQ acknowledges funding from STFC in association with St. CatherineÕs College, Oxford and from HEFCE and STFC through a lectureship funded by the South East Physics Network (SEPnet).


\appendix

\section{Derivation of the free energy expansion near the quantum critical endpoint}

In this Appendix, we derive the expansion Eq.~\ref{eq:GL} from the free energy expression Eq.~\ref{eq:energies} at $T=0$.  While the energy itself is a continuous function, the presence of van-Hove singularities in the density of states leads to logarithmic divergences in derivatives.  The goal is as follows, following the spirit of Landau: to expand the Free energy near the van-Hove energies to obtain the leading order logarithmic terms, and then by using phenomenological arguments in performing the rest of the calculations and adding the different terms while keeping the correct physics, we obtain the expansion Eq.~\ref{eq:GL} in the main text.

The kinetic energy can be written as (we measure energies in units of $t_\|$)
\be
\la F_0 \ra = \int \frac{d^2{\bf k}}{(2\pi)^2} \xi_0({\bf k})n({\bf k}) = g - 2(x-x_0) f
\ee
where
\bea
g &=& \int \frac{d^2{\bf k}}{(2\pi)^2} \xi^*({\bf k})n({\bf k})  \nn
f &=& \int \frac{d^2{\bf k}}{(2\pi)^2} \cos k_y n({\bf k}).
\eea
Some elementary manipulations transform the potential energy into the form
\bea
\la F_1 \ra &=& V \int \frac{d^2{\bf k}}{(2\pi)^2} \int \frac{d^2{\bf q}}{(2\pi)^2} \cos(q_y) n({\bf k}) \left[ 1- n({\bf k}+{\bf q}) \right] \nn
&=& -V f^2.
\eea
We now define the normal density of states
\be
\nu(\xi) = 2\pi \int \frac{d^2{\bf k}}{(2\pi)^2} \delta\left(\xi-\xi^*({\bf k})\right),
\ee
and a weighted density of states
\be
\tilde{\nu}(\xi) = 2\pi \int \frac{d^2{\bf k}}{(2\pi)^2} \cos k_y \, \delta\left(\xi-\xi^*({\bf k})\right).
\ee
This allows us to write our integrals as
\bea
g &=& \int^0_{-\Lambda} \frac{d\xi}{2\pi} \xi \, \nu(\xi) \nn
f &=& \int^0_{-\Lambda} \frac{d\xi} {2\pi} \tilde{\nu}(\xi),
\eea
where $-\Lambda$ is the bottom of the band.  As we will be interested in the derivative of the energy with respect to small changes in the distribution function, the values of such energy integrals at their lower limit is unimportant, and will be ignored in the next.

The density of states $\nu(\xi)$ has a logarithmic divergence at the van-Hove energies $\xi_{\text{vH}}$, which is a result of the integration around the saddle points at $(\pm \pi,0)$ and $(0,\pm \pi)$.  While the energy of these two points coincides in the case $t_\perp=t_\|$, for the anisotropic model we study, they occur at two different energies.  For the case $t_\perp<t_|$ and less than half filling (which is what we deal with in the present manuscript), the important points are $(0,\pm \pi)$.  This gives
\be
\nu(\xi) \propto -\ln(\xi-\xi_{\text{vH}})
\ee
and similarly as $\cos k_y=-1$ at these points,
\be
\tilde{\nu}(\xi) \propto \ln(\xi-\xi_{\text{vH}}).
\ee
Taking into account that the van-Hove energy is proportional to the parameter $x$, it transforms the leading logarithmic terms into
\bea
g &\propto& x^2\ln|x| + \text{regular terms} \nn
f & \propto& x\ln|x|  + \text{regular terms}.
\eea
At this stage, while the regular terms are sub-leading and thus unimportant for small enough $x$, a little care must be taken with them in order to obtain the correct physical behavior.  In particular, the kinetic energy $g-2(x-x_0)f$ whilst showing the above non-analytic behavior as $x\rightarrow 0$ must also have a minimum at $x=x_0$.  The simplest addition of an $x^2$ term ensures this, giving
\be
\la F_0 \ra \sim x^2 \left[ \ln|x| - \frac{1}{2} \right] - 2x\left[ x - x_0 \right] \ln|x| 
\ee
As the potential energy is proportional to $f^2$, one must here also consider the constant term in $f\sim c + x\ln|x|$, which gives the expansion Eq.~\ref{eq:GL} in the main text.



\begin{thebibliography}{10}

\bibitem{Pom58} I. Ia .Pomeranchuk, Sov. Phys. JETP {\bf 35}, 524 (1958).

\bibitem{Yamase-Kohno-2000} H. Yamase and H. Kohno, J. Phys. Soc. Japan {\bf 69}, 2151--2157 (2000).

\bibitem{Halboth-Metzner-2000} C. J. Halboth and W. Metzner, Phys. Rev. Lett. {\bf 85}, 5162--5165 (2000).

\bibitem{Valenzuela-Vozmediano-2001} B. Valenzuela and M. A. H. Vozmediano,  Phys. Rev. B {\bf 63}, 153103 (2001).

\bibitem{Oganesyan-Kivelson-Fradkin-2001} V. Oganesyan, S. A. Kivelson and E. Fradkin, Phys. Rev. B {\bf 64}, 195109 (2001).

\bibitem{Settai-et-al-2005} R. Settai {\it et al.}, J. Phys. Soc. Jpn. {\bf 74}, 3016 (2005).

\bibitem{Sugawara-et-al-2002} H. Sugawara {\it et al.}, Phys. Rev. B {\bf 66}, 134411 (2002).

\bibitem{Paschen-et-al-2004}  S. Paschen {\it et al.}, Nature {\bf 432}, 881 (2004).

\bibitem{Cooper-et-al-2002} K. B. Cooper, M. P. Lilly, J. P. Eisenstein, L. N. Pfeiffer and K. W. West Phys. Rev. B {\bf 65}, 241313(R) (2002).

\bibitem{Grigera-et-al-2004} S. A. Grigera {\it et al.}, Science {\bf 306}, 1154 (2004).

\bibitem{Mercure-et-al-2009} J.-F. Mercure {\it et al.}, Phys. Rev. Lett. {\bf 103}, 176401 (2009).

\bibitem{Quintanilla-Schofield-2006} J. Quintanilla and A. J. Schofield, Phys. Rev. B {\bf 74} 115126 (2006).

\bibitem{Varma-Zhu-2006} C. M. Varma and L. Zhu, Phys. Rev. Lett. {\bf 96} 36405 (2006).

\bibitem{Varma-Zhu-2007} C. M. Varma and L. Zhu, Phys. Rev. Lett. {\bf 98} 177004 (2007).

\bibitem{Doan-Manausakis-2007} Q. M. Doan and E. Manousakis, Phys. Rev. B {\bf 75} 195433 (2007).

\bibitem{Yamase-2009a} H. Yamase, Phys. Rev. Lett. {\bf 102}, 116404 (2009);

\bibitem{Yamase-2009b}  H. Yamase, Phys. Rev. B {\bf 80}, 115102 (2009).

\bibitem{Lee-Wu-2009} Wei-Cheng Lee and Congjun Wu, Phys. Rev. B {\bf 80}, 104438 (2009).

\bibitem{Zacharias-Wolfle-Garst-2009} M. Zacharias, P. Wolfle, and M. Garst, Phys. Rev. B {\bf 80}, 165116 (2009).

\bibitem{Puetter-Rau-Kee-2010} C. M. Puetter, J. G. Rau, and Hae-Young Kee, Phys. Rev. B {\bf 81}, 081105(R) (2010).

\bibitem{Shannon-Momoi-Sindzingre-2006} N. Shannon, T. Momoi and P. Sindzingre, Phys. Rev. Lett. {\bf 96}, 027213 (2006).

\bibitem{Kivelson-Fradkin-Emery-98} S. A. Kivelson, E. Fradkin and V. J. Emery, Nature {\bf 393}, 550 (1998).

\bibitem{Fradkin-Kivelson-Oganesyan-07} E. Fradkin, S. A. Kivelson and V. Oganesyan, Science {\bf 315}, 196 (2007).

\bibitem{Chen-Cheong-1996} C.H. Chen and S.-W. Cheong, Phys. Rev. Lett. {\bf 76}, 4042 (1996).

\bibitem{Tranquada-et-al-1995} J. M. Tranquada, B. J. Sternlieb, J D. Axe, Y. Nakamuara and S. Uchida, Nature {\bf 375}, 561 (1995).

\bibitem{Clarke-et-al-94} D. G. Clarke, S. P. Strong and P. W. Anderson, Phys. Rev. Lett. {\bf 72}, 3218 (1994).

\bibitem{Arrigoni-99} E. Arrigoni, Phys. Rev. Lett. {\bf 83}, 128 (1999).

\bibitem{Arrigoni-00} E. Arrigoni, Phys. Rev. B {\bf 61}, 7909 (2000).

\bibitem{Ledowski-Kopietz-07} S. Ledowski and P. Kopietz, Phys. Rev. B {\bf 76}, 121403 (2007).

\bibitem{Betouras-Chalker} J. J. Betouras and J. T. Chalker,  Phys. Rev. B {\bf 62}, 10931 (2000). 

\bibitem{Campo-et-al-07} V. L. Campo, K. Capelle, J. Quintanilla and C. Hooley, Phys. Rev. Lett. {\bf 99}, 240403 (2007).

\bibitem{Ho-Zhou-2010} Tin-Lun Ho and Qi Zhou, Nat. Phys. {\bf 6}, 131--134 (2010).

\bibitem{2005-Baranov-et-al}M. A. Baranov, L. Dobrek, and M. Lewenstein, Phys. Rev. Lett. {\bf 92}, 250403 (2004).

\bibitem{Buchler-et-al-07} H. P. B\"{u}chler, E. Demler, M. Lukin, A. Micheli, N. Prokof'ev, G. Pupillo and P. Zoller, Phys. Rev. Lett. {\bf 98}, 060404 (2007).

\bibitem{Micheli-et-al-07} A. Micheli, G. Pupillo, H. P. B\"{u}chler and P. Zoller, Phys. Rev. A {\bf 76}, 043604 (2007).

\bibitem{Ni-et-al-2008} K.-K. Ni, S. Ospelkaus, M. H. G. de Miranda,  A. Pe'er, B. Neyenhuis, J. J. Zirbel, S. Kotochigova,  P. S. Julienne, D. S. Jin, and J. Ye, Science {\bf 322}, 231 (2008).

\bibitem{Lu-Youn-Lev-2010} Mingwu Lu, Seo Ho Youn, and Benjamin L. Lev, Phys. Rev. Lett. 104, 063001 (2010). 

\bibitem{2008-Miyakawa-Sogo-Pu} T. Miyakawa, T. Sogo, and H. Pu,  Phys. Rev. A {\bf 77}, 061603(R) (2008).

\bibitem{Quintanilla-Carr-Betouras-2009} J. Quintanilla, S. T. Carr, J. J. Betouras,  Phys. Rev. A {\bf 79}, 031601(R) (2009).

\bibitem{Carr-Quintanilla-Betouras-2009} S. T. Carr, J. Quintanilla, J. J. Betouras, Int. J. Mod. Phys. B  {\bf 23}, 4074 (2009).

\bibitem{Fregoso-et-al-2009} B. M. Fregoso, K. Sun, E. Fradkin and  B. L. Lev, New J. of Physics, {\bf 11}, 103003 (2009).

\bibitem{2009-Cooper-Shlyapnikov} N. R. Cooper, and G. V. Shlyapnikov, Phys. Rev. Lett. {\bf 103}, 155302 (2009).

\bibitem{2009-Huang-Wang} Y.-P. Huang and D.-W. Wang, Phys. Rev. A {\bf 80}, 053610 (2009).

\bibitem{Zhao-et-al-2009} C.Zhao, L.Jiang, X.Liu, W.M.Liu, X.Zou and H.Pu, arXiv:0910.4775 (2009).

\bibitem{Lin-et-al-2009} C.Lin, E.Zhao and W.V.Liu, Phys. Rev. B {\bf 81}, 045115 (2010).

\bibitem{Lifshitz-1960} I. M. Lifshitz, Sov. Phys. JETP {\bf 11}, 1130 (1960).

\bibitem{Perry-et-al-2001} R. S. Perry {\it et al.}, Phys. Rev. Lett. {\bf 86}, 2661 (2001).

\bibitem{Grigera-et-al-2001} S. A. Grigera {\it et al.}, Science {\bf 294}, 329 (2001).

\bibitem{Yamaji-Misawa-Imada-2006} Y. Yamaji, T. Misawa and M. Imada, J. Phys. Soc. Jpn. {\bf 75}, 094719 (2006).

\bibitem{Okamoto-Nishio-Hiroi-2010} Y. Okamoto, A. Nishio, and Z. Hiroi, Phys. Rev. B {\bf 81}, 121102(R) (2010).

\bibitem{Edwards-2006} D. M. Edwards, Physica B {\bf 378-380}, 133--134 (2006)

\bibitem{Yamaji-Misawa-Imada-2007} Y. Yamaji, T. Misawa and M. Imada, J. Phys. Soc. Jpn. {\bf 76}, 063702 (2007). 

\bibitem{Dzyaloshinskii-Larkin-1974} I.E.Dzyaloshinskii and A.I.Larkin, Sov. Phys. JETP, {\bf} 38, 202 (1974)

\bibitem{Solyom-1978} J. Solyom, Advances in Physics {\bf 28}, 201 (1978).

\bibitem{Carr-Tsvelik} S. T. Carr, and A. M. Tsvelik, Phys. Rev. B {\bf 65}, 195121 (2002). 

\bibitem{Gorkov-Dzyaloshinskii} L. P. Gorkov and I. E. Dzyaloshinskii, Zh. Eksp. Teor. Fiz. {\bf 67}, 397 (1974).

\bibitem{Mihaly-Solyom} L. Mihaly and J. Solyom, J. Low Temp. Phys. {\bf 24}, 579 (1976).

\bibitem{Menyhard} N. Menyhard, Solid State Communications {\bf 21}, 495 (1977).

\bibitem{Solyom-1973} J. Solyom, J. Low. Temp. Phys. {\bf 12}, 547 (1973).

\bibitem{Shin-et-al-2008}Yong-il Shin {\it et al.}, Nature {\bf 451}, 689-693 (2008).

\bibitem{Bloch-2005} I. Bloch, Nature Physics, {\bf 1}, 23 (2005).

\bibitem{Kohl-et-al-2005} M. K\"{o}hl, H. Moritz, T. St\"{o}ferle, K. G\"{u}nter and T. Esslinger, Phys. Rev. Lett. {\bf 94}, 080403 (2005).

\bibitem{2001-Valenzuela-Vozmediano} B. Valenzuela, and M. A. H. Vozmediano, Phys. Rev. B {\bf 63},  153103 (2001).

\bibitem{2008-Quintanilla-et-al} J. Quintanilla, C. Hooley, B.J. Powell, A.J. Schofield, and M. Haque. Physica B {\bf 403}, 1279 (2008).

\bibitem{Feldman-Salmhofer} J. Feldman, M. Salmhofer, Rev. Math. Phys. {\bf 20}, 275 (2008).

\bibitem{Bickers-Scalapino-White-1989} N. E. Bickers, D. J. Scalapino and S. R. White, Phys, Rev. Lett. {\bf 62}, 961 (1989).

\bibitem{Chubukov-Efremov-Eremin-2008} A. V. Chubukov, D. V. Efremov and I. Eremin, Phys. Rev. B {\bf 78}, 134512 (2008).

\bibitem{2005-Abbamonte-et-al} P. Abbamonte, A. Rusydi, S. Smadici, G. D. Gu, G. A. Sawatzky, and D. L. Feng, Nature Physics {\bf 1} 155 (2005).

\bibitem{2002-Ando-et-al}Y. Ando, K. Segawa, S. Komiya, A. N. Lavrov, Phys. Rev. Lett. {\bf 88}, 137005 (2002).

\bibitem{2008-Hinkov-et-al} V. Hinkov {\it et al.},  Science {\bf 319}, 597 (2008).

\bibitem{2008-Wise-et-al} W. D. Wise {\it et al.}, Nature Physics {\bf 4},  696 (2008).

\end{thebibliography}
\end{document}